\documentclass[pre,twocolumn]{revtex4-2}
\usepackage{amsmath}
\usepackage{amssymb}
\usepackage{graphicx}
\usepackage{braket}
\usepackage{stmaryrd}
\usepackage{hyperref}
\usepackage{color}
\usepackage{dsfont}
\usepackage{bm}

\hypersetup{
    colorlinks,
    citecolor=blue,
    linkcolor=blue,
    urlcolor=blue
}

\begin{document}

\title{
Dynamics-Independent Bounds on State Transformations and Precision\\ in Open Quantum Systems
}
\author{Yoshihiko Hasegawa}
\email{hasegawa@biom.t.u-tokyo.ac.jp}
\affiliation{Department of Information and Communication Engineering, Graduate
School of Information Science and Technology, The University of Tokyo,
Tokyo 113-8656, Japan}

\date{\today}
\begin{abstract}

We derive dynamics-independent upper bounds on achievable quantum state transformations. 
Modeling the evolution as a joint unitary on the system and its environment, we show that the R\'enyi divergence between the initial system state and any state reachable via the dynamics is bounded from above by a quantity determined solely by the eigenvalues of the initial system and environment density operators.
As a consequence, we establish dynamics-independent lower bounds on the relative variance for arbitrary measurements, which parallel thermodynamic uncertainty relations.
Moreover, we obtain dynamics- and measurement-independent lower bounds on the variance of parameter estimators. These results depend only on the initial eigenvalues of the system and environment and hold for any joint unitary, providing computable bounds for open quantum systems.

\end{abstract}
\maketitle

\begin{figure}
\includegraphics[width=1\linewidth]{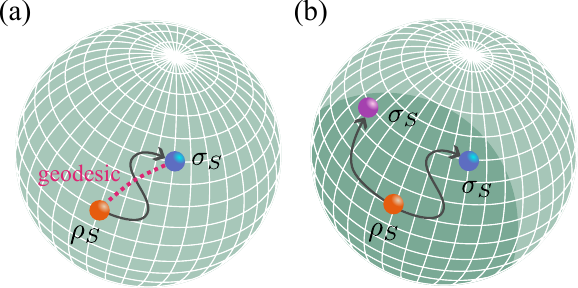}
\caption{
Conceptual illustration of bounds on state transformations. We start with an initial state $\rho_{SE} = \rho_S \otimes \rho_E$, where the joint system can undergo any unitary transformation. (a) Many quantum speed-limit results consider a single, specific time evolution.
They evaluate a distance between the initial state $\rho_S$ and the final state $\sigma_S$ and establish an upper bound on the distance that depends on the dynamics. 
Upper bounds correspond to the distance traveled along the actual time evolution while the distance between the initial state $\rho_S$ and the final state $\sigma_S$ is the geodesic length. 
(b) 
Dynamics-independent bounds considered in this Letter take a broader approach by considering all possible time evolutions from the initial state $\rho_S\otimes \rho_E$.
That is, they apply to essentially any state within the shaded region.
The upper bound on the distance depends on the eigenvalues of $\rho_S$ and $\rho_E$. 
Importantly, these parameters are independent of any specific unitary transformation, making these bounds universal across all possible evolutionary paths.
}
\label{fig:comparison}
\end{figure}

\textit{Introduction.}---Relativity dictates that nothing travels faster than light. Even in nonrelativistic quantum mechanics, there is an upper bound on the speed of state transformation.
This concept is known as the quantum speed limit, which sets a fundamental limit on how quickly a quantum state can change \cite{Mandelstam:1945:QSL,Margolus:1998:QSL,Deffner:2010:GenClausius,Taddei:2013:QSL,DelCampo:2013:OpenQSL,Deffner:2013:DrivenQSL,Pires:2016:GQSL,OConnor:2021:ActionSL} (see \cite{Deffner:2017:QSLReview} for a review). Because the quantum speed limit applies to any system governed by quantum mechanics, it has many applications in quantum computation \cite{Lloyd:2000:CompLimit,Howard:2023:QubitQSL}, quantum communication \cite{Bekenstein:1981:InfoTransfer,Murphy:2010:QSLchain}, and quantum thermodynamics \cite{Deffner:2010:GenClausius} to name but a few. The concept of speed limits has recently been extended to classical systems \cite{Shiraishi:2018:SpeedLimit,Ito:2018:InfoGeo,Ito:2020:TimeTURPRX}. 
More specifically, the quantum speed limit by Mandelstam and Tamm \cite{Mandelstam:1945:QSL} states
\begin{align}
    \mathcal{L}\left(\ket{\psi(0)},\ket{\psi(\tau)}\right)\leq\sqrt{\mathrm{Var}_{\ket{\psi(0)}}[H]}\tau.
    \label{eq:MT_bound}
\end{align}
where $\tau>0$ is the final time, $\mathrm{Var}_{\ket{\psi(0)}}[H]\equiv\braket{\psi(0)|H^{2}|\psi(0)}-\braket{\psi(0)|H|\psi(0)}^{2}$ denotes the variance of $H$ evaluated at the initial state, and $\mathcal{L}(\rho,\sigma)$ denotes the Bures angle defined by
$\mathcal{L}(\rho,\sigma)\equiv\arccos\sqrt{\mathrm{Fid}(\rho,\sigma)}$, 
where $\mathrm{Fid}(\rho,\sigma)\equiv\left[\mathrm{Tr}\sqrt{\sqrt{\rho}\sigma\sqrt{\rho}}\right]^{2}$ is the quantum fidelity.
When considering the orthogonal state for $\ket{\psi(\tau)}$, the left-hand side of Eq.~\eqref{eq:MT_bound} becomes $\pi/2$, which recovers the original expression due to Ref.~\cite{Mandelstam:1945:QSL}. 
The quantum speed limit has been extended to open quantum systems, in which quantum states evolve under the influence of their environment. For such dynamics, the following inequality holds \cite{Taddei:2013:QSL}:
\begin{align}
    \mathcal{L}\left(\rho(0),\rho(\tau)\right)\leq\frac{1}{2}\int_{0}^{\tau}\sqrt{\mathcal{I}(t)}dt,
    \label{eq:QFI_QSL}
\end{align}
where $\mathcal{I}(t)$ is the quantum Fisher information $\mathcal{I}(t)\equiv(8/dt^{2})\left[1-\sqrt{\mathrm{Fid}(\rho(t),\rho(t+dt))}\right]$. 
In essence, the left-hand side is the geodesic distance between the initial and final states, while the right-hand side is the distance traveled along the actual time evolution (Fig.~\ref{fig:comparison}(a)).
The quantum speed limit in Eq.~\eqref{eq:QFI_QSL} serves as a basis for several speed limits \cite{Taddei:2013:QSL,Pires:2016:GQSL,GarciaPintos:2022:OSP,Hasegawa:2023:BulkBoundaryBoundNC,Nishiyama:2024:NonHermiteQSL,Yadin:2024:QSL}. 
Note that quantum speed limits are typically expressed as the minimum time required for a system to evolve. However, they fundamentally describe the maximum rate at which quantum states can transform as shown by Eqs.~\eqref{eq:MT_bound} and \eqref{eq:QFI_QSL} (see the End Matter).

Equation~\eqref{eq:QFI_QSL} generalizes Eq.~\eqref{eq:MT_bound}. In the special case of closed systems with time-independent Hamiltonians, the two become equivalent because the quantum Fisher information reduces to $\mathcal{I}(t)=4\mathrm{Var}_{\ket{\psi(0)}}[H]$, because the variance of the Hamiltonian is a conserved quantity.
However, they differ fundamentally in their interpretation.
The key distinction lies in how their bounds are calculated. Equation~\eqref{eq:MT_bound} depends only on the evolution time $\tau$ and the variance of the Hamiltonian $H$. Since this variance is fully determined by the Hamiltonian and initial state $\ket{\psi(0)}$, the bound can be computed immediately without tracking the system's dynamics.
In contrast, Eq.~\eqref{eq:QFI_QSL} requires knowledge of the full time evolution $\rho(t)$ to evaluate its bound. Even when the initial state $\rho(0)$ and initial Hamiltonian $H(t=0)$ are known, one cannot predict its right-hand side without first solving for the complete dynamics. This dependence on the full time evolution significantly limits its predictive utility.

In this Letter, we take an alternative approach to deriving upper bounds on quantum state transformations.
Instead of considering a state transformation based on a differential equation, 
we consider a joint unitary evolution of the system and environment.
We establish dynamics-independent bounds on achievable state transformations (Fig.~\ref{fig:comparison}(b)). Specifically, we show that the R\'enyi divergence between the initial system state $\rho_S$ and any reachable state $\sigma_S$ is bounded by an expression that depends only on the eigenvalues of initial density operators (cf. Eq.~\eqref{eq:main_result}). As a consequence, we prove dynamics-independent lower bounds on the relative variance for arbitrary measurements (cf. Eq.~\eqref{eq:TUR1}), which parallel thermodynamic uncertainty relations \cite{Barato:2015:UncRel,Gingrich:2016:TUP,Erker:2017:QClockTUR,Brandner:2018:Transport,Carollo:2019:QuantumLDP,Liu:2019:QTUR,Guarnieri:2019:QTURPRR,Saryal:2019:TUR,Hasegawa:2020:QTURPRL,Hasegawa:2020:TUROQS,Kalaee:2021:QTURPRE,Monnai:2022:QTUR,Hasegawa:2023:BulkBoundaryBoundNC,Nishiyama:2024:NonHermiteQSLPRA,Prech:2025:CoherenceQTUR,Moreira:2025:MultiTUR}.
Moreover, we derive measurement- and dynamics-independent lower bounds on the variance of parameter estimators (cf. Eq.~\eqref{eq:Chapman_Robbins_lowerbound}). 
All bounds require only the initial eigenvalues of the system and environment and hold for any joint unitary, thereby providing computable bounds in open quantum systems.

\begin{figure}
\includegraphics[width=0.95\linewidth]{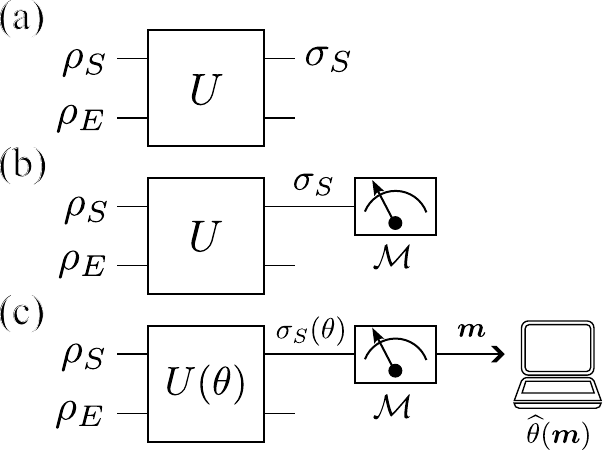}
    \caption{Schematic representation of three scenarios for analyzing quantum processes.
(a) In bounds on state transformations, the focus is on the difference between the initial state $\rho_S$ and the final state $\sigma_S$ after the evolution.
(b) In thermodynamic uncertainty relations, attention is given to the ratio between the variance $\mathrm{Var}_{\sigma_S}[\mathcal{M}]$ and the squared mean $(\mathbb{E}_{\sigma_S}[\mathcal{M}]-\mathbb{E}_{\rho_S}[\mathcal{M}])^2$ of measurement $\mathcal{M}$ performed on the final system state $\sigma_S$.
(c) In estimation theory, the joint unitary $U(\theta)$ contains an unknown parameter $\theta$, and measurements $\mathcal{M}$ on the final system state $\sigma_S(\theta)$.
This process is repeated $R$ times to obtain $\bm{m}=[m_1,m_2,\cdots,m_R]$, which
are used to compute an estimator $\widehat{\theta}_R(\bm{m})$, with emphasis on the variance of the estimator $\mathrm{Var}_\theta[\widehat{\theta}_R(\bm{m})]$.}
    \label{fig:esimation}
\end{figure}

\textit{Preliminaries.}---
In this Letter, we consider open quantum systems.  
We consider a principal system $S$ and an environment $E$, whose initial states are denoted by $\rho_S$ and $\rho_E$, respectively. 
Here, the dimensions of $S$ and $E$ are defined as $d_S$ and $d_E$, respectively. 
The initial state of $S+E$ is given by $\rho_S \otimes \rho_E$,  
and $\rho_S \otimes \rho_E$ evolves over time under a joint unitary operator $U$.  
The state of the principal system after time evolution is given by the following (Fig.~\ref{fig:esimation}(a)):
\begin{align}
    \sigma_{S}=\mathrm{Tr}_{E}\left[U\left(\rho_{S}\otimes\rho_{E}\right)U^{\dagger}\right],
    \label{eq:rhoS_prime_def}
\end{align}
where the partial trace over the environment is denoted by $\mathrm{Tr}_E$.
We are interested in the upper bound on the distance between $\rho_S$ and $\sigma_S$. 
Here, we focus on the R\'enyi relative entropy, which generalizes the quantum relative entropy and subsumes several important divergences as special cases.
Let us introduce the divergences for the general density operators $\rho$ and $\sigma$. 
For a parameter $\alpha \in (0,1)\cup(1,\infty)$, the Petz-R\'enyi relative entropy is defined by \cite{Petz:1986:QuasiEntropies}
\begin{align}
    D_{\alpha}(\rho\|\sigma)\equiv\frac{1}{\alpha-1}\ln\mathrm{Tr}\left[\rho^{\alpha}\sigma^{1-\alpha}\right].
    \label{eq:Petz_Renyi_relative_entropy}
\end{align}
Another generalization of the quantum relative entropy is the sandwiched R\'enyi relative entropy
\cite{MullerLennert:2013:RenyiEntropy,Leditzky:2016:PhD}:
\begin{align}
    \widetilde{D}_{\alpha}(\rho\|\sigma)\equiv\frac{1}{\alpha-1}\ln\left[\mathrm{Tr}\left[\left(\sigma^{\frac{1-\alpha}{2\alpha}}\rho\sigma^{\frac{1-\alpha}{2\alpha}}\right)^{\alpha}\right]\right],
    \label{eq:Sandwitched_Renyi_relative_entropy}
\end{align}
which is also defined for $\alpha \in (0,1)\cup(1,\infty)$. 
Both the Petz and sandwiched variants reduce to the quantum relative entropy as $\alpha \to 1$:
\begin{align}
    \lim_{\alpha\to1}D_{\alpha}(\rho\|\sigma)=\lim_{\alpha\to1}\widetilde{D}_{\alpha}(\rho\|\sigma)=\mathrm{Tr}\left[\rho(\ln\rho-\ln\sigma)\right].
    \label{eq:Renyi_to_relative_entropy}
\end{align}
Moreover, the sandwiched R\'enyi relative entropy reduces to the quantum fidelity for $\alpha = 1/2$:
\begin{align}
    \mathrm{Fid}(\rho,\sigma)=e^{-\widetilde{D}_{1/2}(\rho\|\sigma)}.
    \label{eq:SRRE_fidelity}
\end{align}
In the classical limit, both reduce to the R\'enyi divergence:
\begin{align}
    \mathfrak{D}_{\alpha}(P\|Q)\equiv\frac{1}{\alpha-1}\ln\left[\sum_{x}P(x)^{\alpha}Q(x)^{1-\alpha}\right].
    \label{eq:classical_Renyi_div_def}
\end{align}
Here, $P(x)$ and $Q(x)$ are probability distributions. 
Let $a_n$ and $b_n$ be the eigenvalues of $\rho$ and $\sigma$, respectively, where the dimension of $\rho$ and $\sigma$ is $d$.
Throughout, we use $d$ for the dimension in general discussions; for specific cases, we use context-appropriate symbols.
For any vector $\mathbf{x} \in \mathbb{R}^d$, we define $x^\uparrow$ as the vector obtained by sorting the components of $\mathbf{x}$ in non-decreasing order. That is, $x_1^\uparrow \leq x_2^\uparrow \leq \cdots \leq x_d^\uparrow$, where $x_n^\uparrow$ denotes the $n$-th smallest element of $x$. Similarly, $x^\downarrow$ represents the vector sorted in non-increasing order, satisfying $x_1^\downarrow \geq x_2^\downarrow \geq \cdots \geq x_d^\downarrow$, where $x_n^\downarrow$ is the $n$-th largest element of $\mathbf{x}$.
Using the von Neumann trace inequality (see the End Matter), it is known that
\begin{align}
    D_{\alpha}(\rho\|\sigma)\leq\frac{1}{\alpha-1}\ln\left[\sum_{n=1}^{d}\left(a_{n}^{\uparrow}\right)^{\alpha}\left(b_{n}^{\downarrow}\right)^{1-\alpha}\right].
    \label{eq:Petz_RE_upperbound}
\end{align}
Regarding the magnitude relation between Eqs.~\eqref{eq:Petz_Renyi_relative_entropy} and \eqref{eq:Sandwitched_Renyi_relative_entropy},
it is known that  \cite{Wilde:2014:RenyiEntropy,Datta:2014:RenyiDiv,MullerLennert:2013:RenyiEntropy}
\begin{align}
    D_\alpha(\rho\|\sigma)\ge \widetilde{D}_\alpha(\rho\|\sigma).
    \label{eq:D_and_Dtilde_relation}
\end{align}
Therefore, for the sandwiched R\'enyi relative entropy, the same upper bound holds:
\begin{align}
    \widetilde{D}_{\alpha}(\rho\|\sigma)\leq\frac{1}{\alpha-1}\ln\left[\sum_{n=1}^{d}\left(a_{n}^{\uparrow}\right)^{\alpha}\left(b_{n}^{\downarrow}\right)^{1-\alpha}\right].
    \label{eq:sandwitch_RE_upperbound}
\end{align}

\textit{Results.}---
Let $A$ be an Hermitian operator and $\lambda(A)$ be a set of eigenvalues of $A$. 
The notation $\lambda^\downarrow_n(A)$ refers to the $n$-th largest eigenvalue of the Hermitian operator $A$, where the eigenvalues are arranged in non-increasing order,
that is,
$\lambda_{1}^{\downarrow}\ge\lambda_{2}^{\downarrow}\ge\cdots$
($\lambda^\uparrow_n(A)$ is defined analogously).
As mentioned above, we are interested in upper bounds on the divergence between $\rho_S$ and $\sigma_S$. 
From Eq.~\eqref{eq:Petz_RE_upperbound}, we have
\begin{align}
    {D}_{\alpha}(\rho_{S}\|\sigma_{S})\leq\frac{1}{\alpha-1}\ln\left[\sum_{n=1}^{d_{S}}\lambda_{n}^{\uparrow}(\rho_{S})^{\alpha}\lambda_{n}^{\downarrow}(\sigma_{S})^{1-\alpha}\right],
    \label{eq:sandwiched_rhoS_sigmaS_upperbound}
\end{align}
which depend on $\lambda(\rho_S)$ and $\lambda(\sigma_S)$ (note that $d=d_S$ for this case). 
From Eq.~\eqref{eq:D_and_Dtilde_relation}, the same upper bound holds for $\widetilde{D}_\alpha(\rho_{S}\|\sigma_{S})$. 
The central observation is that the eigenvalues $\lambda{(\sigma_S)}$ cannot take arbitrary values
given the initial density operators $\rho_S$ and $\rho_E$. 
Since the joint state obeys unitary transformation, 
sets of 
eigenvalues before and after the unitary evolution agree:
\begin{align}
    \lambda(U(\rho_{S}\otimes\rho_{E})U^{\dagger})=\lambda(\rho_{S}\otimes\rho_{E}).
    \label{eq:eigenvalue_equiv}
\end{align}
Moreover, the eigenvalues of a tensor product state are given by the products of the eigenvalues of each factor:
\begin{align}
\Lambda \equiv \lambda(\rho_{S}\otimes\rho_{E}) = \{ab \mid a \in \lambda(\rho_{S}), \; b \in \lambda(\rho_{E})\},
\label{eq:rhoS_rhoE_eigenvalues}
\end{align}
where $|\Lambda| = d_S d_E$. 
Therefore, the eigenvalues of states after a unitary is $\lambda(U(\rho_S\otimes \rho_E)U^\dagger) = \Lambda$. 
Let $\sigma_S$ and $\sigma_S^\prime$ be density operators after the evolution. 
Suppose the eigenvalue vector of $\sigma_S$ majorizes that of $\sigma_S^\prime$. 
That is $\lambda(\sigma_{S})\succ\lambda(\sigma_{S}^{\prime})$,
where we identify sets as vectors.
Then, by the Schur-Ostrowski criterion, 
the right-hand side of Eqs.~\eqref{eq:Petz_RE_upperbound} and \eqref{eq:sandwitch_RE_upperbound} is bounded from above by (see the End Matter)
\begin{align}
&\frac{1}{\alpha-1}\ln\!\left[\sum_{n=1}^{d_{S}}\lambda_{n}^{\uparrow}(\rho_{S})^{\alpha}\lambda_{n}^{\downarrow}(\sigma_{S})^{1-\alpha}\right]\nonumber\\&\ge\frac{1}{\alpha-1}\ln\!\left[\sum_{n=1}^{d_{S}}\lambda_{n}^{\uparrow}(\rho_{S})^{\alpha}\lambda_{n}^{\downarrow}(\sigma_{S}^{\prime})^{1-\alpha}\right],
\label{eq:upperbound_majorization}
\end{align}
which holds for $\alpha >0$ except $\alpha = 1$. 
Therefore, Eq.~\eqref{eq:sandwiched_rhoS_sigmaS_upperbound} reduces the task to finding the eigenvalue vector $\lambda(\sigma_S)$ that majorizes all other eigenvalue vectors obtainable via the open quantum dynamics in Eq.~\eqref{eq:rhoS_prime_def}.
This problem is reminiscent of thermodynamic cooling \cite{Allahverdyan:2011:Cooling,Clivaz:2019:Cooling}.
Since $\lambda(\sigma_S)$ is obtained by taking the partial trace over the environment $E$,
we can compute it explicitly when $U(\rho_S \otimes \rho_E)U^\dagger$ is given in diagonal form:
the partial trace corresponds to summing the diagonal entries within each $d_E \times d_E$ block of the matrix.
The set of eigenvalues of $U(\rho_S \otimes \rho_E)U^\dagger$ is $\Lambda$ from Eq. \eqref{eq:rhoS_rhoE_eigenvalues}.
Hence, an optimal $\sigma_S^*$ is obtained by taking the partial trace of a diagonal matrix whose entries are arranged in decreasing order as $[\Lambda_1^\downarrow, \Lambda_2^\downarrow, \dots, \Lambda_{d_S d_E}^\downarrow]$ \cite{Clivaz:2019:Cooling},
where the notation $\Lambda^\downarrow_n$ refers to the $n$-th largest value in $\Lambda$.
We have restricted our attention to the diagonal case.
However, the Schur-Horn theorem tells us that for an Hermitian matrix, the set of all possible diagonal vectors (in some basis) is the set of vectors that are majorized by its eigenvalue vector.
This means that when we search for $\sigma_S$ under a majorization constraint, we can restrict our attention to diagonal matrices.

Let us define
\begin{align}
C_{k}=\sum_{n=1}^{kd_{E}}\Lambda_{n}^{\downarrow}.
    \label{eq:Sigma_k_def}
\end{align}
Let $\sigma_S^*$ be the optimal post dynamics state 
under the majorization constraint. 
Note that the largest eigenvalue of $\lambda(\sigma_S^*)$ is $C_1$. 
The second largest eigenvalue is $C_2 - C_1$.
The third largest eigenvalue is obtained in a similar manner. 
Therefore,
the eigenvalues of the optimal $\sigma^*_S$ are
\begin{align}
    \lambda(\sigma_{S}^*)=[C_{1},C_{2}-C_{1},\cdots,C_{d_{S}}-C_{d_{S}-1}],
    \label{eq:sigmaS_maj_eigenvalues}
\end{align}
By construction, the eigenvalue vector of $\sigma_S^*$ majorizes the eigenvalue vectors of all states achievable via Eq.~\eqref{eq:rhoS_prime_def}. 
Using Eqs.~\eqref{eq:upperbound_majorization} and \eqref{eq:sigmaS_maj_eigenvalues}, we obtain
\begin{align}
    {D}_{\alpha}(\rho_{S}\|\sigma_{S})\leq\frac{1}{\alpha-1}\ln\left[\sum_{n=1}^{d_{S}}\lambda_{n}^{\uparrow}(\rho_{S})^{\alpha}(C_{n}-C_{n-1})^{1-\alpha}\right],
    \label{eq:main_result}
\end{align}
where $C_0 \equiv 0$ for notational convenience. 
Since $C_n$ depends only on $\lambda(\rho_S)$ and $\lambda(\rho_E)$, 
the upper bound in Eq.~\eqref{eq:main_result} depends only on the eigenvalues of $\rho_S$ and $\rho_E$. 
Equation~\eqref{eq:main_result} is the first result of this Letter.

The R\'enyi relative entropy given by Eqs.~\eqref{eq:Petz_Renyi_relative_entropy} and \eqref{eq:Sandwitched_Renyi_relative_entropy} can cover several important statistical measures. 
As given in Eq.~\eqref{eq:SRRE_fidelity}, for $\alpha=1/2$, 
the sandwiched R\'enyi relative entropy reduces to the fidelity. 
Using this upper bound, the Bures angle, which plays a central role in the quantum speed limit, is bounded from above by
\begin{align}
\mathcal{L}(\rho_{S},\sigma_{S})\le\arccos\left[\sum_{n=1}^{d_{S}}\sqrt{\lambda_{n}^{\uparrow}(\rho_{S})(C_{n}-C_{n-1})}\right].
    \label{eq:Bures_angle_bound}
\end{align}
Equation~\eqref{eq:Bures_angle_bound} shows that there exists a fundamental upper bound, independent of the dynamics, on the Bures angle between the initial and reachable system states.

The upper bound on quantum state distances has important implications for measurement processes. To illustrate this, consider performing a measurement with a positive operator-valued measure (POVM) in state $\sigma_S$ following a unitary interaction (see Fig.~\ref{fig:esimation}(b)).
Let $P(x)$ and $Q(x)$ be probability distributions. 
The $\chi^2$ divergence is defined by
\begin{align}
    \chi^{2}(P\|Q)\equiv\sum_{x}\frac{(P(x)-Q(x))^{2}}{Q(x)}.
    \label{eq:chi2_div_def}
\end{align}
When considering continuous distributions, the summation should be replaced by the integration. 
Let $g(x)$ be a real function.
Then, from the variational representation of the $\chi^2$ divergence,
it is well known that the following inequality holds \cite{Rockafellar:1970:ConvexBook}:
\begin{align}
    \chi^{2}(P\|Q)\ge\frac{\left(\mathbb{E}_{P}[g(X)]-\mathbb{E}_{Q}[g(X)]\right)^{2}}{\mathrm{Var}_{Q}[g(X)]},
    \label{eq:chi2_variational_repr}
\end{align}
where $\mathbb{E}_{P}[g(X)]\equiv\sum_{x}P(x)g(x)$ and $\mathrm{Var}_{Q}[g(X)]\equiv\mathbb{E}_{Q}[g(X)^{2}]-\mathbb{E}_{Q}[g(X)]^{2}$. 
A key implication of Eq.~\eqref{eq:chi2_variational_repr} is that the $\chi^2$-divergence is bounded from below by the mean and variance of an arbitrary function $g(X)$.
Equation~\eqref{eq:chi2_variational_repr} directly leads to 
Hammersley-Chapman-Robbins \cite{Hammersley:1950:HCRBound,Chapman:1951:HCRBound} inequality, which generalizes
the Cram\'er-Rao inequality \cite{Cramer:1946:Book}. 

Consider a POVM $\mathcal{M} = \{M_m\}_{m \in \mathcal{O}}$, where $\mathcal{O}$ is the set of possible real-valued measurement outcomes.
 POVM elements satisfy $\sum_{m\in\mathcal{O}}M_{m}=\mathbb{I}$ and $M_m \ge 0$ (positive semi-definite). 
Let $P(m|\rho)\equiv \mathrm{Tr}[M_m \rho]$ denote the probability of measuring $m\in\mathcal{O}$ given the state $\rho$. 
Let $\mathbb{E}_{\rho}[\mathcal{M}]\equiv\sum_{m}mP(m|\rho)$ and $\mathrm{Var}_{\rho}[\mathcal{M}]\equiv\sum_{m}m^{2}P(m|\rho)-\mathbb{E}_{\rho}[\mathcal{M}]^{2}$,
which are the mean and variance of POVM $\mathcal{M}$ evaluated with the density operator $\rho$. 
Using the variational representation given in Eq.~\eqref{eq:chi2_variational_repr}, we obtain
\begin{align}
    \chi^{2}(P(m|\rho_{S})\|P(m|\sigma_{S}))\ge\frac{\left(\mathbb{E}_{\rho_{S}}[\mathcal{M}]-\mathbb{E}_{\sigma_{S}}[\mathcal{M}]\right)^{2}}{\mathrm{Var}_{\sigma_{S}}[\mathcal{M}]}.
    \label{eq:chisquare_TUR}
\end{align}
The R\'enyi relative entropy with $\alpha=2$ and the $\chi^2$-divergence is related via
\begin{align}
    {D}_{2}(\rho_{S}\|\sigma_{S})&\ge\mathfrak{D}_{2}(P(m|\rho_{S})\|P(m|\sigma_{S}))\nonumber\\&=\ln\left[1+\chi^{2}(P(m|\rho_{S})\|P(m|\sigma_{S}))\right].
    \label{eq:D2_and_chisquare}
\end{align}
The first inequality in Eq.~\eqref{eq:D2_and_chisquare} is due to the monotonicity relation of the Petz R\'enyi entropy under a CPTP for $\alpha \in (0,1)\cup(1,2]$ \cite{Petz:1986:QuasiEntropies}. 
Using Eqs.~\eqref{eq:main_result}, \eqref{eq:chisquare_TUR}, and \eqref{eq:D2_and_chisquare}, we obtain
\begin{align}
    \frac{\mathrm{Var}_{\sigma_{S}}[\mathcal{M}]}{\left(\mathbb{E}_{\sigma_{S}}[\mathcal{M}]-\mathbb{E}_{\rho_{S}}[\mathcal{M}]\right)^{2}}\ge\left[\sum_{n=1}^{d_{S}}\frac{\lambda_{n}^{\uparrow}(\rho_{S})^{2}}{C_{n}-C_{n-1}}-1\right]^{-1},
    \label{eq:TUR1}
\end{align}
which is the second result of this Letter. 
The variance is evaluated with respect to the state after unitary evolution. The mean in the denominator is the mean value after unitary evolution, with the initial state used as the reference. As in Eq.~\eqref{eq:main_result}, the right-hand side depends only on the initial states of $S$ and $E$.
Equation~\eqref{eq:TUR1} is a reminiscent of thermodynamic uncertainty relations,
where the relative variance, the variance divided by the squared mean, is bounded from below by a measurement-independent term. 
Given the initial state, Eq.~\eqref{eq:TUR1} constitutes the precision limit in open quantum systems, which cannot be surpassed by any choice of joint unitary.
A similar dynamics-independent bound for the relative variance was recently derived in Ref.~\cite{Hasegawa:2024:PrecisionLimit}.
Because Ref.\cite{Hasegawa:2024:PrecisionLimit} relies on the probability that the smallest eigenvalue is realized, it does not apply to continuous distributions. In contrast, Eq.~\eqref{eq:TUR1} remains valid for continuous distributions.

In the third scenario, we focus on parameter estimation in open quantum systems (Fig.~\ref{fig:esimation}(c)). 
We prepare a probe state $\rho_S$ that is to undergo a dynamical process $\mathcal{K}_\theta$ (completely positive and trace preserving map) parametrized by an unknown parameter $\theta \in \mathbb{R}$. 
Here, we assume that the probe undergoes an open quantum dynamics described by (Fig.~\ref{fig:esimation}(c))
\begin{align}
    \sigma_{S}(\theta)&=\mathrm{Tr}_{E}\left[U(\theta)(\rho_{S}\otimes\rho_{E})U^{\dagger}(\theta)\right]\nonumber\\&=\mathcal{K}_{\theta}(\rho_{S}).
    \label{eq:sigmaS_theta_def}
\end{align}
It is natural to assume that $\sigma_S(\theta_0)=\rho_S$,
which implies that the parametrized dynamics $\mathcal{K}_\theta$ includes the identity operation. 
For example, when $U(\theta)=e^{-i G \theta}$, where $G$ is a generator operator, $\sigma_S(\theta_0)=\rho_S$ with $\theta_0 = 0$. 
The objective is to estimate the parameter $\theta \in \mathbb{R}$.
After the time evolution $\sigma_S(\theta) = \mathcal{K}_\theta(\rho_S)$, we apply measurement operator $\mathcal{M} = \{M_m\}_{m\in\mathcal{O}}$ to obtain a measurement outcome $m$.
We repeat this process $R$ times. 
The probe is prepared in the tensor product state $\Upsilon_{S}(\theta) = \bigotimes_{n=1}^{R}\rho_{S}^{(n)}(\theta)$, where each component $\rho_{S}^{(n)}(\theta)$ undergoes the transformation $\mathcal{K}_\theta$. This process yields the output state $\Sigma_{S}(\theta) = \bigotimes_{n=1}^{R}\sigma_{S}^{(n)}(\theta)$.
We then apply the measurement operator $\mathcal{M}$ to each state in
$\Sigma_S(\theta)$ and obtain the measurement outcomes $\bm{m} \equiv [m_1,m_2,\cdots,m_R]$. 
Finally, we process the measurement outcomes $\boldsymbol{m}$ to construct an estimator $\widehat{\theta}_R(\bm{m})$ of the unknown parameter $\theta$.

Let us define the expectation of $\widehat{\theta}_R(\bm{m})$:
\begin{align}
    \mathbb{E}_{\theta}\left[h\left(\widehat{\theta}_{R}\right)\right]\equiv\sum_{\bm{m}}P(\bm{m}|\Sigma_{S}(\theta))h\left(\widehat{\theta}_{R}(\bm{m})\right),
    \label{eq:Expectation_thetahat_R}
\end{align}
where
$h(\bullet)$ is an arbitrary real function. 
When we have $R$ independent measurements and the estimator is unbiased (i.e., $\mathbb{E}_\theta[\widehat{\theta}_R] = \theta$), 
the quantum Cram\'er-Rao bound holds \cite{Helstrom:1967:Estimation,Paris:2009:QFI,Liu:2019:QTUR}:
\begin{align}
    \mathrm{Var}_\theta[\widehat{\theta}_R]\ge\frac{1}{R\mathcal{I}(\theta)},
    \label{eq:quantum_CRB}
\end{align}
where $\mathrm{Var}_{\theta}[\widehat{\theta}_{R}]\equiv\mathbb{E}_{\theta}[\widehat{\theta}_{R}^{2}]-\mathbb{E}_{\theta}[\widehat{\theta}_{R}]^{2}=\mathbb{E}_{\theta}[\widehat{\theta}_{R}^{2}]-\theta^{2}$ and $\mathcal{I}(\theta)$ denotes the quantum Fisher information. 
However, analytical treatment of the quantum Fisher information in mixed states of open quantum systems is notoriously difficult \cite{Paris:2009:QFI}.
In this Letter, we provide an analytical lower bound on the variance of $\widehat{\theta}_R$ in open quantum systems by combining Eq.~\eqref{eq:chi2_variational_repr} with Eq.~\eqref{eq:main_result}.
Using the additivity of the R\'enyi relative entropy, given a tensor product states, we have
\begin{align}
    {D}_{\alpha}\left(\Upsilon_{S}\middle\|\Sigma_{S}(\theta)\right)=R{D}_{\alpha}\left(\rho_{S}\|\sigma_{S}(\theta)\right).
    \label{eq:Dalpha_additive}
\end{align}
Using Eqs.~\eqref{eq:main_result}, \eqref{eq:D2_and_chisquare} and \eqref{eq:chi2_variational_repr},
we obtain (see the End Matter).
\begin{align}
    \frac{\mathrm{Var}_{\theta}[\widehat{\theta}_{R}]}{\left(\mathbb{E}_{\theta}[\widehat{\theta}_{R}]-\mathbb{E}_{\theta_{0}}[\widehat{\theta}_{R}]\right)^{2}}\ge\left[\left(\sum_{n=1}^{d_{S}}\frac{\lambda_{n}^{\uparrow}(\rho_{S})^{2}}{C_{n}-C_{n-1}}\right)^{R}-1\right]^{-1}.
    \label{eq:Chapman_Robbins_lowerbound}
\end{align}
Note that $\widehat{\theta}_R$ in Eq.~\eqref{eq:Chapman_Robbins_lowerbound} need not be unbiased.
Equation~\eqref{eq:Chapman_Robbins_lowerbound} is the third result of this Letter. 
From Eq.~\eqref{eq:Chapman_Robbins_lowerbound}, we can obtain the lower bound for the mean squared error of $\widehat{\theta}_R$. 
We define the mean squared error (MSE) of $\widehat{\theta}_{R}$ as $\mathrm{MSE}\left[\widehat{\theta}_{R}\right]=\mathbb{E}[(\widehat{\theta}_{R}-\theta)^{2}]$, and the squared bias as $\mathrm{Bias}\left[\widehat{\theta}_{R}\right]=\left(\mathbb{E}[\widehat{\theta}_{R}]-\theta\right)^{2}$.
In statistics, the mean squared error evaluates the accuracy of an estimator. It represents the average of the squared differences between the estimated value $\widehat{\theta}_R$ and the true value $\theta$. 
Since $\mathrm{MSE}\left[\widehat{\theta}_{R}\right]=\mathrm{Var}\left[\widehat{\theta}_{R}\right]+\mathrm{Bias}\left[\hat{\theta}_{R}\right]$,
we obtain the lower bound on $\mathrm{MSE}[\widehat{\theta}_R]$:
\begin{align}
    \frac{\mathrm{MSE}(\widehat{\theta}_{R})}{\left(\mathbb{E}_{\theta}[\widehat{\theta}_{R}]-\mathbb{E}_{\theta_{0}}[\widehat{\theta}_{R}]\right)^{2}}\ge\left[\left(\sum_{n=1}^{d_{S}}\frac{\lambda_{n}^{\uparrow}(\rho_{S})^{2}}{C_{n}-C_{n-1}}\right)^{R}-1\right]^{-1}.
    \label{eq:MSE_bound}
\end{align}
Equations~\eqref{eq:Chapman_Robbins_lowerbound} and \eqref{eq:MSE_bound} provide lower bounds on the variance of the estimator and the mean squared error, respectively.  
In the quantum Cram{\'e}r-Rao inequality, the quantum Fisher information depends on the parameter being estimated.  
On the other hand, the lower bound given by Eqs.~\eqref{eq:Chapman_Robbins_lowerbound} and \eqref{eq:MSE_bound} only requires condition $\sigma_S(\theta_0) = \rho_S$ at $\theta = \theta_0$.  
Furthermore, although it is generally difficult to handle the quantum Fisher information analytically, the right-hand side of Eqs.~\eqref{eq:Chapman_Robbins_lowerbound} and \eqref{eq:MSE_bound} has the advantage that it depends only on the eigenvalues of the initial states $\rho_S$ and $\rho_E$.

Finally, we comment on the effects of coherence, i.e., off-diagonal elements in density operators. 
Since Eq.~\eqref{eq:main_result} depends only on eigenvalues, it is basis independent and therefore cannot directly capture coherence effects.
Here, we compare the case of a density operator with only diagonal elements to that of a non-diagonal density operator that shares the same diagonal elements.
According to the Schur-Horn theorem, the eigenvalues of a Hermitian matrix majorize its diagonal elements. 
Therefore, when considering $\rho_S$, the presence of off-diagonal elements always affects the right-hand side of Eq.~\eqref{eq:main_result}. Similarly, for $\rho_E$, off-diagonal elements have the same effect on the right-hand side of Eq.~\eqref{eq:main_result}.

\textit{Conclusion.}---We established dynamics-independent bounds on achievable state transformations in open quantum systems by treating the joint system-environment evolution as a joint unitary. Our main result upper bounds the Petz and sandwiched R\'enyi divergences between an initial system state and any reachable state using only the eigenvalues of the initial system and environment. 
We further showed that these bounds provide operational constraints for measurements and estimation. For arbitrary POVMs, we derived a dynamics- and measurement-independent lower bound on the relative variance, which parallels thermodynamic uncertainty relations. For parameter estimation, we derived lower bounds on estimator variance and mean-squared error that bypass the technical challenges of the quantum Fisher information.

\begin{acknowledgments}

This work was supported by JSPS KAKENHI Grant Numbers JP23K24915 and JP24K03008.

\end{acknowledgments}

\appendix

\section*{End Matter}

\section{Quantum speed limit}

Quantum speed limits set the fastest possible rate at which quantum systems can change. Although often presented as minimum time requirements for evolution, they ultimately stem from fundamental inequalities governing how quickly quantum states can transform.
Let us define the time-averaged quantity:
\begin{align}
    \left\langle \int_{0}^{\tau}\sqrt{\mathcal{I}(t)}dt\right\rangle _{\tau}\equiv\frac{1}{\tau}\int_{0}^{\tau}\sqrt{\mathcal{I}(t)}dt.
    \label{eq:time_everaged_Ft}
\end{align}
Using Eq.~\eqref{eq:time_everaged_Ft}, Eq.~\eqref{eq:QFI_QSL} can be rewritten as
\begin{align}
    \mathcal{L}\left(\rho(0),\rho(\tau)\right)\leq\frac{\tau}{2}\left\langle \int_{0}^{\tau}\sqrt{\mathcal{I}(t)}dt\right\rangle _{\tau},
    \label{eq:QSL_1_EM}
\end{align}
which leads to
\begin{align}
    \tau\ge\frac{2\mathcal{L}\left(\rho(0),\rho(\tau)\right)}{\left\langle \int_{0}^{\tau}\sqrt{\mathcal{I}(t)}dt\right\rangle _{\tau}}.
    \label{eq:QSL_2_EM}
\end{align}
Equation~\eqref{eq:QSL_2_EM} provides the lower bound for the time required to transition from $\rho(0)$ to $\rho(\tau)$.

\section{Derivation of Eq.~\eqref{eq:upperbound_majorization}}

A Schur-concave function is a symmetric function $f: \mathbb{R}^d \rightarrow \mathbb{R}$ such that $\mathbf{x} \prec \mathbf{y}$ implies $f(\mathbf{x}) \geq f(\mathbf{y})$.
The Schur-Ostrowski criterion provides a test for determining whether a given function $f$ is Schur-concave. According to this criterion, if $f$ is symmetric and has first partial derivatives, then $f$ is Schur-concave if and only if the following condition holds:
\begin{align}
    (x_{n}-x_{m})\left(\frac{\partial f}{\partial x_{n}}-\frac{\partial f}{\partial x_{m}}\right)\le0\,\text{for}\,x_{n}\ne x_{m}.
    \label{eq:Schur_Ostrowski_criterion}
\end{align}
To show Eq.~\eqref{eq:upperbound_majorization}, we consider the following function:
\begin{align}
    f(\mathbf{x})=\sum_{n=1}^{d}(a_{n}^{\uparrow})^{\alpha}(x_{n}^{\downarrow})^{1-\alpha}.
    \label{eq:fx_def_EM}
\end{align}
First, we consider the case $0<\alpha<1$. 
Then
\begin{align}
    &(x_{n}^{\downarrow}-x_{m}^{\downarrow})\left(\frac{\partial f}{\partial x_{n}^{\downarrow}}-\frac{\partial f}{\partial x_{m}^{\downarrow}}\right)\nonumber\\&=(1-\alpha)(x_{n}^{\downarrow}-x_{m}^{\downarrow})\left((a_{n}^{\uparrow})^{\alpha}(x_{n}^{\downarrow})^{-\alpha}-(a_{m}^{\uparrow})^{\alpha}(x_{m}^{\downarrow})^{-\alpha}\right)\nonumber\\&\le0.
    \label{eq:SO_criterion_1}
\end{align}
Equation~\eqref{eq:SO_criterion_1} shows that $f(\mathbf{x})$ defined in Eq.~\eqref{eq:fx_def_EM} is Schur-concave. Therefore, given $\mathbf{b},\mathbf{c}\in \mathbb{R}^{d}$ satisfying $\mathbf{b} \succ \mathbf{c}$, we have
\begin{align}
    \sum_{n=1}^{d}(a_{n}^{\uparrow})^{\alpha}(b_{n}^{\downarrow})^{1-\alpha}\le\sum_{n=1}^{d}(a_{n}^{\uparrow})^{\alpha}(c_{n}^{\downarrow})^{1-\alpha}\,(0<\alpha<1).
    \label{eq:condition_1_EM}
\end{align}
In a similar way, we can show that, for $\alpha > 1$
\begin{align}
    \sum_{n=1}^{d}(a_{n}^{\uparrow})^{\alpha}(b_{n}^{\downarrow})^{1-\alpha}\ge\sum_{n=1}^{d}(a_{n}^{\uparrow})^{\alpha}(c_{n}^{\downarrow})^{1-\alpha}\,(\alpha>1).
    \label{eq:condition_2_EM}
\end{align}
Equations~\eqref{eq:condition_1_EM} and \eqref{eq:condition_2_EM} prove 
\begin{align}
\frac{1}{\alpha-1}\ln\!\left[\sum_{n=1}^{d}(a_{n}^{\uparrow})^{\alpha}(b_{n}^{\downarrow})^{1-\alpha}\right]\ge\frac{1}{\alpha-1}\ln\!\left[\sum_{n=1}^{d}(a_{n}^{\uparrow})^{\alpha}(c_{n}^{\downarrow})^{1-\alpha}\right].
\label{eq:upperbound_majorization_EM}
\end{align}

\section{Upper bound of Petz R\'enyi relative entropy}

We derive Eq.~\eqref{eq:Petz_RE_upperbound}.
Let $A$ and $B$ be $d$-dimensional positive semi-definite matrices.
The von Neumann inequality states
\begin{align}
    \sum_{n=1}^{d}\lambda_{n}^{\uparrow}(A)\lambda_{n}^{\downarrow}(B)\le\mathrm{Tr}[AB]\le\sum_{n=1}^{d}\lambda_{n}^{\uparrow}(A)\lambda_{n}^{\uparrow}(B).
    \label{eq:von_Neumann_ineq}
\end{align}
The eigenvalues of $A^\alpha$ is $\lambda_n(A)^\alpha$. 
For $\alpha > 1$, the following relation holds:
\begin{align}
    \mathrm{Tr}[\rho^{\alpha}\sigma^{1-\alpha}]&\le\sum_{n=1}^{d}\lambda_{n}^{\uparrow}(\rho^{\alpha})\lambda_{n}^{\uparrow}(\sigma^{1-\alpha})\nonumber\\&=\sum_{n=1}^{d}(a_{n}^{\uparrow})^{\alpha}(b_{n}^{\downarrow})^{1-\alpha}.
    \label{eq:UB_alpha_larger_1}
\end{align}
Similarly, for $0<\alpha<1$, the following relation holds:
\begin{align}
    \mathrm{Tr}[\rho^{\alpha}\sigma^{1-\alpha}]&\ge\sum_{n=1}^{d}\lambda_{n}^{\uparrow}(\rho^{\alpha})\lambda_{n}^{\downarrow}(\sigma^{1-\alpha})\nonumber\\&=\sum_{n=1}^{d}(a_{n}^{\uparrow})^{\alpha}(b_{n}^{\downarrow})^{1-\alpha}.
    \label{eq:LB_alpha_zero_one}
\end{align}
Using Eqs.~\eqref{eq:UB_alpha_larger_1} to \eqref{eq:LB_alpha_zero_one}, we obtain
Eq.~\eqref{eq:Petz_RE_upperbound}.

\section{Estimator variance bound}

Here we prove the estimator variance bound shown in Eq.~\eqref{eq:Chapman_Robbins_lowerbound}. 
From Eqs.~\eqref{eq:D2_and_chisquare} and \eqref{eq:chisquare_TUR}, we have
\begin{align}
    &D_{2}\left(\Upsilon_{S}\middle\|\Sigma_{S}(\theta)\right)\nonumber\\&\ge\ln\left[1+\chi^{2}\left(P(\bm{m}|\Upsilon_{S})\middle\|P(\bm{m}|\Sigma_{S}(\theta))\right)\right]\nonumber\\&\ge\ln\left[1+\frac{\left(\mathbb{E}_{P(\bm{m}|\Sigma_{S}(\theta))}[g(\bm{m})]-\mathbb{E}_{P(\bm{m}|\Upsilon_{S})}[g(\bm{m})]\right)^{2}}{\mathrm{Var}_{P(\bm{m}|\Sigma_{S}(\theta))}[g(\bm{m})]}\right].
    \label{eq:estimation_ineq_EM}
\end{align}
Recall that
\begin{align}
    \mathbb{E}_{P(\bm{m}|\Sigma_{S}(\theta))}[g(\bm{m})]=\mathbb{E}_{\theta}[g(\bm{m})].
    \label{eq:E_equiv_EM}
\end{align}
which follows from Eq.~\eqref{eq:Expectation_thetahat_R}. 
Using the additivity of the R\'enyi relative entropy [Eq.~\eqref{eq:Dalpha_additive}] and taking $g(\bm{m}) = \widehat{\theta}_R(\bm{m})$ in Eq.~\eqref{eq:estimation_ineq_EM} (recall that $g(\bullet)$ can be an arbitrary function), we obtain
\begin{align}
    \frac{1}{e^{R{D}_{2}(\rho_{S}\|\sigma_{S}(\theta))}-1}\le\frac{\mathrm{Var}_{\theta}[\hat{\theta}_{R}]}{\left(\mathbb{E}_{\theta}[\hat{\theta}_{R}]-\mathbb{E}_{\theta_{0}}[\hat{\theta}_{R}]\right)^{2}}.
    \label{eq:Var_intermediate}
\end{align}
Moreover, 
from Eq.~\eqref{eq:main_result}, we have
\begin{align}
    {D}_{2}(\rho_{S}\|\sigma_{S}(\theta))\leq\ln\left[\sum_{n=1}^{d_{S}}\frac{\lambda_{n}^{\uparrow}(\rho_{S})^{2}}{C_{n}-C_{n-1}}\right].
    \label{eq:D2_upperbound}
\end{align}
Combining Eq.~\eqref{eq:Var_intermediate} with Eq.~\eqref{eq:D2_upperbound}, we obtain Eq.~\eqref{eq:Chapman_Robbins_lowerbound} in the main text.

\end{document}